\documentclass[prl,aps,epsf]{revtex4}
\usepackage[english]{babel}
\usepackage{graphicx}
\usepackage{epsf}
\usepackage{pstricks}

\begin{document}

\hfill{CPHT-RR-037.0703}

\title{On $\gamma\gamma^*$ production of two $\rho^0$ mesons}

\author{I.V. Anikin $^{a,b}$, B. Pire$^a$ and O.V. Teryaev$^{b,c}$}
\affiliation{$^a$CPHT de l'{\'E}cole Polytechnique,
             91128 Palaiseau Cedex, France\footnote{Unit{\'e} Mixte de
             Recherche du CNRS (UMR 7644)} \\
             $^b$Bogoliubov Laboratory of Theoretical Physics,
             JINR, 141980 Dubna, Russia \\
             $^c$CPT-CNRS-Luminy, 13288 Marseille Cedex 9, France
             \footnote {Unit{\'e} Propre de Recherche du
             CNRS (UPR 7061)}}
\vspace{1.5cm}

\begin{abstract}

\noindent
We present a theoretical estimation for the
cross-section of exclusive two $\rho$-meson production in
two photon collision when one of the initial
photons is highly virtual. The compatibility of our analysis
with recent experimental data
obtained by the L3 Collaboration at LEP is discussed. We show that
these data prove the scaling behaviour of the exclusive production amplitude. 
They are thus consistent with a partonic description
of the exclusive
process $\gamma\gamma ^*\to\rho ^0\rho ^0$  when
$Q^2\sim 2 - 20\, {\rm GeV^2}$.
\vspace{1pc}
\end{abstract}
\maketitle

\section{Introduction}

\noindent
Two-photon collisions provide a tool
to study a variety of fundamental aspects of QCD and have long been a
subject of great interest (cf., e.g., \cite{Terazawa,Bud,photon-conf}
and references therein). A peculiar facet of this interesting domain is
exclusive two hadron production
in the region where one initial photon is highly virtual (its
virtuality being denoted as $Q^2$) but the overall energy (or invariant
mass of the two hadrons) is small~\cite{DGPT}. This
process factorizes~\cite{MRG,Freund} into a perturbatively calculable,
short-distance dominated scattering $\gamma^* \gamma \to q \bar q$ or
$\gamma^* \gamma \to g g$, and non-perturbative matrix elements
measuring the transitions $q \bar q \to A B$ and $g g \to  A B $. These
matrix elements  have been called generalized distribution
amplitudes (GDAs) to emphasize their close connection to the
distribution amplitudes introduced many years ago in the QCD
description of exclusive hard processes~\cite{LepageBrodsky}.

\noindent
In this paper, we focus on the process $\gamma^* \gamma \to \rho^0 \rho^0$
which has not been much discussed theoretically (see however Ref.
\cite{Maul}) but has recently been observed at LEP in the right
kinematical domain \cite{L3Coll}.

\noindent
Since  $\gamma^* \gamma \to \rho \rho$ is the
crossed channel of virtual Compton scattering on a spin-$1$ meson, the
physics considered here is closely related to deeply
virtual Compton scattering (DVCS) on a spin-$1$ target~\cite{CP},
which has recently attracted  some attention in the context of generalized
(skewed) parton distributions of the deuteron~\cite{BCDP}.

\section{ Kinematics }

\noindent
The reaction which we study here is (see, Fig. 1):
\begin{eqnarray}
\label{pr}
e(k)+e(l)\to e(k^{\prime})+e(l^{\prime})+\rho ^0(p_1)+\rho ^0(p_2)
\end{eqnarray}
where the initial  electron $e(k)$ radiates
a hard virtual photon with momentum $q=k-k^{\prime}$,
in other words, the square of virtual photon momentum $q^2=-Q^2$ is
very large. This means that the scattered electron  $e(k^{\prime})$
is tagged.
To describe reaction (\ref{pr}), it is useful to consider, at the same
time,  the sub-process :
\begin{eqnarray}
\label{spr}
e(k)+\gamma(q^{\prime})\to e(k^{\prime})+\rho ^0(p_1)+\rho ^0(p_2).
\end{eqnarray}
Regarding the other photon momentum $q^{\prime}=l-l^{\prime}$,
we assume that, firstly, its momentum is collinear to the electron
momentum $l$ and, secondly, that $q^{\prime \, 2}$ is approximately
equal to zero, which is a usual approximation when the second lepton
is untagged.

\noindent
Let us now pass to a short discussion of kinematics in
the $\gamma\gamma ^*$ center of mass system. We adopt the
$z$ axis directed along the three-dimensional vector ${\bf q}$,
and the $\rho$-meson momenta lie in the $(x,z)$-plane.
It means that we ignore the azimuthal dependence of the cross-section, which
happens to be absent in the approximation we use.
So, we write for the momenta in the c.m. system~:
\begin{eqnarray}
\label{kin}
q=(q_0,\,0,\,0,\,{\bf q}),
\quad p_1=(p_1^0,\, {\bf p}_1{\rm sin}\theta,\,0,\,
{\bf p}_1{\rm cos}\theta).
\end{eqnarray}
Also, we need to write down the Mandelstam $S$-variables for
the electron-positron (\ref{pr}) and electron-photon (\ref{spr})
collisions:
\begin{eqnarray}
S_{ee}=(k+l)^2, \quad S_{e\gamma}=(k+q^{\prime})^2.
\end{eqnarray}
Neglecting the lepton masses, these variables
can be rewritten as
\begin{eqnarray}
\label{See}
S_{ee}\approx 2(k\cdot l), \quad
S_{e\gamma}\approx 2(k\cdot q^{\prime})=x_2 S_{ee},
\end{eqnarray}
where the fraction $x_2$ defined as  $q_0^{\prime}=x_2 l_0 $ is introduced
(see, \cite{Diehl00}).
\begin{figure}[htb]
\vspace{9pt}
%\framebox[55mm]{\rule[-21mm]{0mm}{43mm}}
\includegraphics[width=20pc]{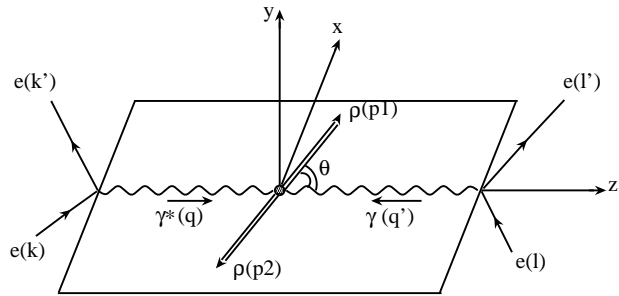}
\vspace{-2cm}
\caption{Kinematics of the process $e(k)+e(l)\to
e(k^{\prime})+e(l^{\prime})+\rho ^0(p_1)+\rho ^0(p_2)$ in the c.m.s of
the two mesons.}
\end{figure}

\section{Parameterization of $\rho$-matrix elements and
their properties }

\noindent
Let us first introduce the basis light-cone vectors.
We adopt a  basis
consisting of two light-like vectors $p$ and $n$
of mass dimension $1$ and $-1$, respectively.
In other words, they obey the following conditions~:
\begin{eqnarray}
\label{lcb}
p^2=n^2=0, \quad (p\cdot n)=1.
\end{eqnarray}
With the help of this basis,
the $\rho$-mesons momenta $p_1 $ and $p_2 $ can be written as
\begin{eqnarray}
\label{lcd1}
&&p_1^{\mu}=\zeta p^{\mu}+
(1-\zeta) \frac{W^2}{2} n^{\mu}-
\frac{\Delta_{T}^{\mu}}{2},
\nonumber\\
&&p_2^{\mu}=(1-\zeta) p^{\mu}+
\zeta \frac{W^2}{2} n^{\mu}+
\frac{\Delta_{T}^{\mu}}{2}
\end{eqnarray}
As usually for the two meson generalized distribution amplitude
case, we introduce the sum and difference of hadronic momenta which
take the form in the light-cone decomposition~:
\begin{eqnarray}
\label{lcd2}
\Delta^{\mu}=p_2^{\mu}-p_1^{\mu}, \quad
P^{\mu}=p_2^{\mu}+p_1^{\mu}.
\end{eqnarray}
The skewedness parameter $\zeta$ is defined by
\begin{eqnarray}
\zeta =\frac{p_1^+}{P^+}=\frac{1+\beta\cos{\theta}}{2},
\quad \beta=\sqrt{1-\frac{4m_{\rho}^2}{W^2}}.
\end{eqnarray}
Note also that within the c.m. frame the transverse component of the
transfer momentum $\Delta_T=(0,{\bf \Delta}_T,0)$ is given by
\begin{eqnarray}
\Delta_T^2=-{\bf \Delta}_T^2=\biggl(4m_{\rho}^2-W^2 \biggr)
\sin^2{\theta}.
\end{eqnarray}
Let us now write down the decomposition  for the
longitudinal and transverse  $\rho$-meson polarization vectors:
\begin{eqnarray}
\label{lcd3}
&&e_{1\, \mu}^{(0)}=\frac{1}{m_{\rho}}\biggl(
p_{1\, \mu}-\frac{m^{2}_{\rho}}{\zeta}n_{\mu}
\biggr),
\nonumber\\
&&e_{1\, \mu}^{(1)}=\frac{2}{\sqrt{{\bf \Delta}_T^2}}\Biggl(
\zeta p_{2\, \mu}-(1-\zeta)p_{1\, \mu}-
\frac{\zeta (2\zeta-1) W^2+{\bf \Delta}_T^2}{2\zeta}n_{\mu}
\Biggr) ,
\nonumber\\
&&e_{1\, \mu}^{(2)}=-\frac{2}{\sqrt{{\bf \Delta}_T^2}}
\varepsilon_{\mu\alpha\beta\gamma}p_{2\, \alpha}p_{1\, \beta}
n_{\gamma},
\end{eqnarray}
for the $\rho$-meson with momentum $p_1$, and
\begin{eqnarray}
\label{lcd3_2}
&&e_{2\, \mu}^{(0)}=\frac{1}{m_{\rho}}\biggl(
p_{2\, \mu}-\frac{m^{2}_{\rho}}{1-\zeta}n_{\mu}
\biggr),
\nonumber\\
&&e_{2\, \mu}^{(1)}=-\frac{2}{\sqrt{{\bf \Delta}_T^2}}\Biggl(
(1-\zeta)p_{1\, \mu}-\zeta p_{2\, \mu}+
\frac{(1-\zeta)(2\zeta-1) W^2-{\bf \Delta}_T^2}{2(1-\zeta)}n_{\mu}
\Biggr) ,
\nonumber\\
&&e_{2\, \mu}^{(2)}=\frac{2}{\sqrt{{\bf \Delta}_T^2}}
\varepsilon_{\mu\alpha\beta\gamma}p_{2\, \alpha}p_{1\, \beta}
n_{\gamma},
\end{eqnarray}
for the other $\rho$-meson with  momentum $p_2$.
As usual,
\begin{eqnarray}
\label{orth}
e_i\cdot p_i=0,\quad
e_i^{(\lambda)}\cdot e_i^{(\lambda^{\prime})}=
-\delta^{\lambda\lambda^{\prime}}.
\end{eqnarray}
Besides, for each meson, the following polarization vectors can be
introduced to define the  light-cone helicity~:
\begin{eqnarray}
e(0)=e^{(0)}, \quad e(\pm)=\frac{\mp e^{(1)}-i e^{(2)}}{\sqrt{2}}.
\end{eqnarray}

\noindent
We now come to the parameterization
of the relevant matrix elements.  Keeping
the terms of leading  twist $2$, the vector
and axial correlators can be written as:
\begin{eqnarray}
\label{par1}
&&\langle p_1,\lambda_1;p_2,\lambda_2 |
\bar\psi(0)\gamma_{\mu} \psi(\lambda n)
| 0 \rangle
\stackrel{{\cal F}}{=}
p_{\mu}\sum_{i}\, e_1^{\alpha}e_2^{\beta}
V^{(i)}_{\alpha\beta}(p_1,p_2,n)
H^{\rho\rho,\,V}_i(y,\zeta,W^2),
\\
\label{par2}
&&\langle p_1,\lambda_1;p_2,\lambda_2 |
\bar\psi(0)\gamma_{\mu} \gamma_{5} \psi(\lambda n)
| 0 \rangle
\stackrel{{\cal F}}{=}
p_{\mu}\sum_{i}\, e_1^{\alpha}e_2^{\beta}
A^{(i)}_{\alpha\beta}(p_1,p_2,n)
H^{\rho\rho,\,A}_i(y,\zeta,W^2).
\end{eqnarray}
Here $\lambda_1$ and $\lambda_2$ are the helicities of $\rho$ mesons
and $\stackrel{{\cal F}}{=}$
denotes the Fourier transformation
with measure ($z_1=\lambda n, z_2=0$) \cite{Ani}:
\begin{eqnarray}
d\mu(y)=dy \,e^{ -iy\,pz_1+i(1-y)\,pz_2}.
\end{eqnarray}
With the help of parity invariance we can show that the vector
tensors $V^{(i)}_{\alpha\beta}$ may be written in terms of five
tensor structures while the axial tensors $A^{(i)}_{\alpha\beta}$
are linear combinations of  four independent structures.
In complete analogy with the analysis of the deuteron generalized
parton distributions \cite{BCDP} we write
\begin{eqnarray}
\label{parBCDP_V}
&&\sum_{i} e_1^{\alpha}e_2^{\beta} V^{(i)}_{\alpha\beta}(p_1,p_2,n)
H^{\rho\rho,\,V}_i(y) =
- (e_1\cdot e_2)\, H^{\rho\rho,\,V}_1(y)+
\nonumber\\
&&\biggl((e_1\cdot n) (e_2\cdot p_1)+(e_2\cdot n) (e_1\cdot p_2)\biggr)\,
H^{\rho\rho,\,V}_2(y) -
\frac{(e_1\cdot p_2)(e_2\cdot p_1)}{2 m^2_{\rho}}\,
H^{\rho\rho,\,V}_3(y)+
\nonumber\\
&&\biggl((e_1\cdot n) (e_2\cdot p_1)-
(e_2\cdot n) (e_1 .p_2)\biggr)\, H^{\rho\rho,\,V}_4(y) +
\biggl(
4 m_{\rho}^2\, (e_1\cdot n)(e_2\cdot n)
+\frac{1}{3} (e_1\cdot e_2) \biggr) H^{\rho\rho,\,V}_5(y)
\end{eqnarray}
for the vector tensor structures and
\begin{eqnarray}
\label{parBCDP_A}
&&\sum_{i} e_1^{\alpha}e_2^{\beta} A^{(i)}_{\alpha\beta}(p_1,p_2,n)
H^{\rho\rho,\,A}_i(y)=
- i \epsilon_{\alpha\beta}^T e^T_{1\,\alpha}
e^T_{2\,\beta}\, H^{\rho\rho,\,A}_1(y)
+
\nonumber\\
&&i \epsilon_{\alpha\beta}^T \Delta_{\alpha}\,
\frac{e_{1\,\beta} (e_2\cdot p_1)+e_{2\,\beta} (e_1\cdot p_2)}{m_{\rho}^2}
H^{\rho\rho,\,A}_2(y)+
i \epsilon_{\alpha\beta}^T \Delta_{\alpha}\,
\frac{e_{1\,\beta} (e_2\cdot p_1)-e_{2\,\beta} (e_1\cdot p_2)}{m_{\rho}^2}
H^{\rho\rho,\,A}_3(y)+
\nonumber\\
&& i \epsilon_{\alpha\beta}^T \Delta_{\alpha}\,
\biggl(e_{1\,\beta} (e_2\cdot n)+e_{2\,\beta} (e_1\cdot n)\biggr)\,
H^{\rho\rho,\,A}_4(y)
\end{eqnarray}
for the axial tensor structures. In (\ref{parBCDP_V}) and
(\ref{parBCDP_A}),  the standard notation  $\epsilon^T_{\alpha\beta}=
\epsilon_{\alpha\beta\gamma\delta}\,p_{\gamma}n_{\delta}$ was used
and the dependence of parameterizing functions (GDA) on $\zeta$ and
$W^2$ is implied.

\noindent
Let us now turn to the consideration of symmetry properties.
The $\gamma\gamma^*$ subprocess is selecting the parts with the
following symmetry:
\begin{eqnarray}
\label{Cprop}
H^{\rho\rho,\,V}_i(y)=-H^{\rho\rho,\,V}_i(1-y),
\quad
H^{\rho\rho,\,A}_i(y)=H^{\rho\rho,\,A}_i(1-y)
\end{eqnarray}

\noindent
Note that the scale dependence of generalized distribution amplitudes
acquired in the
process of factorization of the scattering amplitude has been studied in
\cite{Diehl00} and there are no essential differences between
the $\pi \pi$ channel discussed
there and vector contributions to our $\rho \rho$ case. At the same time,
the evolution of axial contributions are similar to the case of the
distribution amplitudes of singlet axial mesons.
The $W^2$ behaviour of $2\pi$ GDA's has been recently explored
in terms of an impact representation of the hadronization process
$q \bar q \to M M$ \cite{PS}.

\section{Quark-hadron helicity amplitudes}

\noindent
Although the main objects of our investigation are the
generalized distribution amplitudes (GDA) let us begin
from the discussion of
the helicity amplitudes related to the generalized parton
distribution (GPD), which are related to GDAs
by means of $s \leftrightarrow t $ crossing symmetry
\cite{Pol, Radon}. One considers
the quark helicity conserving distributions parameterizing
the combination of the
vector and axial matrix elements~:
\begin{eqnarray}
\label{hme}
{\cal A}_{(\lambda_2 \pm\,~;\, \lambda_1 \pm)}=\frac{1}{2}
\langle p_2,\lambda_2 |
\bar\psi(0)\gamma^+(1 \pm \gamma_5)\psi(z)
|  p_1,\lambda_1 \rangle,
\end{eqnarray}
where $\lambda$ and $\mu=\pm$ denote the helicities of
initial (final) hadrons and quarks, respectively. The matrix elements in
(\ref{hme}) are taken between two $\rho$ mesons.
In the forward limit when $p_1=p_2$  helicity conservation takes
place and requires $\lambda_1+\mu_1=\lambda_2+\mu_2$.
The parity transformation, $\lambda_i \to -\lambda_i$ {\it etc.},
invariance leads to nine independent helicity amplitudes (\ref{hme}).
At the same time the time reversal transformation,
$(\mu_1,\lambda_1) \leftrightarrow (\mu_2,\lambda_2)$,
invariance does not lead to any reduction of the number of
independent structures owing to $\zeta$-dependence (see, for
instance, \cite{BCDP}).

\noindent
As noted before, the crossing transformation relates
the helicity amplitudes referring to
the GDAs to the corresponding
GPDs helicity amplitudes. Indeed, the crossing transformations
imply that the initial hadron is replaced by a final hadron
with the opposite momentum :  $p_1\to -p_1$ and, therefore,
opposite helicities $\lambda_1\to -\lambda_1$. Similarly, we implement
the crossing  replacement for the quark fields. Namely, the final
quark with helicity $\mu_2$ is replaced by the initial antiquark
with helicity $-\mu_2$.
Thus, the quark helicity non-flip amplitudes in
$t$-channel come to the amplitudes in
$s$-channel where the initial quark and anti-quark helicities are
opposite,  and vice versa~:
\begin{eqnarray}
\label{cr}
{\cal A}_{(\lambda_2 \pm \,;\, \lambda_1 \pm)}^{(t)}
\rightarrow  A_{(\lambda_2 -\lambda_1 \,;\, \mp \pm)}^{(s)}.
\end{eqnarray}
Note that the first two indices labeling the helicity amplitudes
in (\ref{cr}) or (\ref{hme})
correspond to the helicities in the final state and the
last two indices -- to the initial state.

\noindent
Let us now focus on the amplitudes in $s$-channel.
As the helicity and chirality of antiquarks are
distinguished by sign (in this paper, we consider the case
of massless quarks),  the chirality conserving quark-antiquark
operator will give the combination of quark and
antiquark fields with  opposite helicities~:
$\bar v_{+}\gamma^+ u_{+} \leftrightarrow
\bar v_{(-)}\gamma^+ u_{(+)}$,
where the bracketed subscripts denote the helicity while
the unbracketed ones denote the chirality.
Therefore, using (\ref{par1}) and (\ref{par2}), we write, forgetting from
now on the $(s)$ subscript,
\begin{eqnarray}
\label{hme2}
A_{(\lambda_2\lambda_1 \,;\,\pm \mp)}=\frac{1}{2}
\langle p_2,\lambda_2 ; p_1,\lambda_1 |
\bar\psi(0)\gamma^+(1 \pm \gamma_5)\psi(z)
|0 \rangle.
\end{eqnarray}

\noindent
Further,
a straightforward calculation derives the expressions
for the helicity amplitudes (cf., \cite{BCDP})~:
\begin{eqnarray}
\label{allhel}
2 A_{(++,+-)}&=&H_1^{\rho\rho,\,V}+B H_3^{\rho\rho,\,V}-
\frac{1}{3} H_5^{\rho\rho,\,V}+H_1^{\rho\rho,\,A}+
2B\biggl[ H_2^{\rho\rho,\,A}+(2\zeta-1) H_3^{\rho\rho,\,A}\biggr],
\nonumber\\
2 A_{(0+,+-)}&=&\sqrt{\frac{2B\zeta}{1-\zeta}}\Biggl(
H_1^{\rho\rho,\,V}- \zeta
\biggl[ H_2^{\rho\rho,\,V}+H_4^{\rho\rho,\,V}\biggr]+
\biggl[ \frac{(2\zeta-1) W^2}{4m^2_{\rho}}+
2(1-\zeta)\biggr]H_3^{\rho\rho,\,V}-
\Biggr.
\nonumber\\
\Biggl.
&&\frac{1}{3}H_5^{\rho\rho,\,V}+
(1-\zeta) H_1^{\rho\rho,\,A}+(1-\zeta)
\biggl[ \frac{(2\zeta-1) W^2}{m^2_{\rho}}+4(1-\zeta)B\biggr]
\biggl[ H_2^{\rho\rho,\,A}-H_3^{\rho\rho,\,A} \biggr]+
\Biggr.
\nonumber\\
\Biggl.
&&\zeta(1-\zeta) H_4^{\rho\rho,\,A}
\Biggr),
\nonumber\\
2 A_{(-+,+-)}&=&-2B\Biggl( \frac{1}{2} H_2^{\rho\rho,\,V}-
(2\zeta-1) H_2^{\rho\rho,\,A}- H_3^{\rho\rho,\,A} \Biggr),
\nonumber\\
2 A_{(00,+-)}&=&\biggl[\left(\frac{1}{2}-\zeta(1-\zeta)\right)
\frac{W^2}{m^2_{\rho}}-
4\left(1-\zeta(1-\zeta)\right)B \biggr]
\biggl[H_1^{\rho\rho,\,V}-\frac{1}{3} H_5^{\rho\rho,\,V} \biggr]-
\nonumber\\
&&\frac{(2\zeta-1) W^2}{2m^2_{\rho}}
\biggl[H_4^{\rho\rho,\,V}+(2\zeta-1) H_2^{\rho\rho,\,V} \biggr]+
2B \biggl[H_2^{\rho\rho,\,V}+(2\zeta-1) H_4^{\rho\rho,\,V} \biggr]+
\nonumber\\
&&\biggl[ \frac{(2\zeta-1)^2 W^4}{8m^4_{\rho}}-
2B\left( \frac{(2\zeta-1)^2 W^2}{2m^2_{\rho}}+
4\zeta(1-\zeta)B \right) \biggr]
H_3^{\rho\rho,\,V}+4\zeta(1-\zeta)H_5^{\rho\rho,\,V}.
\end{eqnarray}
In (\ref{allhel}) the following notation has been used:
\begin{eqnarray}
B=\frac{{\bf \Delta}_T^2}{16m^2_{\rho}\zeta(1-\zeta)}.
\end{eqnarray}

\section{Amplitude of $\gamma\gamma^*\to\rho^0\rho^0$ subprocess}

In this section, we  consider the
$\gamma(q^\prime)\gamma^*(q)\to\rho^0(p_1)\rho^0(p_2)$ subprocess.
Following \cite{Ani}, the amplitude of this subprocess including
the leading twist-$2$ terms can be written as
\begin{eqnarray}
\label{amp_gg}
T_{\mu\nu}^{\gamma\gamma^*\to\rho^0\rho^0} =
\frac{1}{2}\sum_{q=1}^{n_f}e_{q}^{2}\int\limits_{0}^{1}
dy\Biggl[
g_{\mu\nu}^T E_-(y)
{\bf V}_q(y,\cos{\theta},W^2) -
i \epsilon^T_{\mu\nu} E_+(y) {\bf A}_q(y,\cos{\theta},W^2)
\Biggr],
\end{eqnarray}
where
\begin{eqnarray}
E_\pm=\frac{1}{1-y}\pm\frac{1}{y}.
\end{eqnarray}
In (\ref{amp_gg}), the scalar and pseudo-scalar
functions (${\bf V}$, ${\bf A}$) denote the following contractions
\begin{eqnarray}
&&{\bf V}_q(y,\cos{\theta},W^2)=
\sum_i e_1^{\alpha}e_2^{\beta}  V^{(i)}_{\alpha\beta}
H^{\rho\rho,\,V}_{i,\,q}(y,\zeta(\cos{\theta}),W^2),
\nonumber\\
&&{\bf A}_q(y,\cos{\theta},W^2)=
\sum_i e_1^{\alpha}e_2^{\beta} A^{(i)}_{\alpha\beta}
H^{\rho\rho,\,A}_{i,\,q}(y,\zeta(\cos{\theta}),W^2).
\nonumber
\end{eqnarray}

\noindent
The helicity amplitudes 
are obtained from the usual amplitudes after  multiplying by
the photon polarization vectors
\begin{eqnarray}
\label{hel}
A_{(i,j)}=\varepsilon^{\,\prime\,(i)}_{\mu} \varepsilon^{(j)}_{\nu}
T^{\mu\nu}_{\gamma\gamma^*\to\rho^0\rho^0}.
\end{eqnarray}
Here, in the $\gamma\gamma ^*$ c.m. frame,
the photon polarization vectors read~
\begin{eqnarray}
\label{pol_vec}
&&\varepsilon^{\,\prime\,(\pm)}_{\mu}=
\left( 0,\frac{\mp 1}{\sqrt{2}},\frac{+i}{\sqrt{2}},0 \right),
\nonumber\\
&&\varepsilon^{\,(\pm)}_{\mu}=
\left( 0,\frac{\mp 1}{\sqrt{2}},\frac{-i}{\sqrt{2}},0 \right),
\quad
\varepsilon^{\,(0)}_{\mu}=
\left(\frac{|q|}{\sqrt{Q^2}},0,0,\frac{q_0}{\sqrt{Q^2}} \right),
\end{eqnarray}
for the real and virtual photons, respectively.

\section{Differential cross sections}

\noindent
We will now concentrate on the calculation of the differential
cross section of (\ref{pr}).
The amplitude of this  process can be written as
\begin{eqnarray}
\label{amp_pr}
{\cal A}_{ee\to ee\rho^0\rho^0}=\sum\limits_{i,j}
\biggl[ \bar u(l^{\prime})\gamma^{\mu}u(l)
\stackrel{*}{\varepsilon}^{\, \prime \,(i)}_{\mu} \biggr]
\frac{1}{q^{\prime\, 2}} A_{(i,j)}^{\gamma\gamma^*\to \rho^0\rho^0}
\frac{1}{q^{2}}
\biggl[\stackrel{*}{\varepsilon}^{(j)}_{\nu}
 \bar u(k^{\prime})\gamma^{\nu}u(k) \biggr].
\end{eqnarray}
This amplitude depends on the polarization states of the produced
$\rho$ mesons.
Due to parity invariance there are only three independent sets of
helicity (photon) amplitudes which we put to be $A_{(+,+)}$, $A_{(+,-)}$ and
$A_{(+,0)}$. Let us focus on  the leading twist-$2$ helicity amplitude
in the unpolarized electrons case, i.e. the $A_{(+,+)}$ amplitude.
In this case, the square of the modulus of the amplitude (\ref{amp_pr})
can be presented in the "factorized" form :
\begin{eqnarray}
\left| {\cal A}_{ee\to ee\rho^0\rho^0}\right|^2=
\left| {\cal A}_{e\gamma\to e\rho^0\rho^0}\right|^2
\frac{1}{q^{\prime\, 4}}
\left| {\cal A}_{e\to e\gamma}\right|^2 
\end{eqnarray}

\noindent
and the scattering cross section is :
\begin{eqnarray}
\label{xsec1}
&&d\sigma_{ee\to ee\rho^0\rho^0}=
\frac{1}{2 S_{ee}}
\frac{d^3\, l^{\prime}}{(2\pi)^3 2 l^{\prime}_0} \,
\frac{d^3\, p_1}{(2\pi)^3 2 p_1^0}\,
\frac{d^3\, p_2}{(2\pi)^3 2 p_2^0}\,
\frac{d^3\, k^{\prime}}{(2\pi)^3 2 k^{\prime}_0}\,
\nonumber\\
&&\left| {\cal A}_{e\gamma\to e\rho^0\rho^0}\right|^2
\frac{1}{q^{\prime\, 4}}
\left| {\cal A}_{e\to e\gamma}\right|^2.
\end{eqnarray}
>From now in, we will sum over the polarization states of the $\rho$ mesons.
Separating the differential cross section for
$e\gamma\to e\rho^0\rho^0$ subprocess, we are able
to rewrite (\ref{xsec1}) in the form :
\begin{eqnarray}
\label{xsec2}
d\sigma_{ee\to ee\rho^0\rho^0}=
\frac{d^3\, l^{\prime}}{(2\pi)^3 2 l^{\prime}_0} \,
\frac{x_2}{q^{\prime\, 4}}
\left| {\cal A}_{e\to e\gamma}\right|^2 \,
d\sigma_{e\gamma\to e\rho^0\rho^0},
\end{eqnarray}
where
\begin{eqnarray}
\label{xsec3}
d\sigma_{e\gamma\to e\rho^0\rho^0}=
\frac{1}{2S_{e\gamma}}\,
\frac{d^3\, p_1}{(2\pi)^3 2 p_1^0}\,
\frac{d^3\, p_2}{(2\pi)^3 2 p_2^0}\,
\frac{d^3\, k^{\prime}}{(2\pi)^3 2 k^{\prime}_0}\,
\left| {\cal A}_{e\gamma\to e\rho^0\rho^0}\right|^2.
\end{eqnarray}

\noindent
Using the equivalent photon approximation
we find the expression for the corresponding cross section~:
\begin{eqnarray}
\label{xsec5}
\frac{d\sigma_{ee\to ee\rho^0\rho^0}}{dQ^2}=
\int..\int dW^2\, d{\rm cos}\theta\, d\phi\, dx_2
\frac{\alpha}{\pi} F_{WW}(x_2)
\frac{d\sigma_{e\gamma\to e\rho^0\rho^0}}
{ dQ^2 \,dW^2\, d{\rm cos}\theta\, d\phi},
\end{eqnarray}
where the Weizsacker-Williams function $F_{WW}$ is defined
as usual as~:
\begin{eqnarray}
\label{FWW}
F_{WW}(x_2)=\frac{1+(1-x_2)^2}{2x_{2}}
{\rm ln}\frac{Q^{\prime\,2}(x_2)}{m_e^2}
-\frac{1-x_2}{x_{2}},
\end{eqnarray}
and the value $Q^{\prime\, 2}$ is defined as
\begin{eqnarray}
\label{Qpr}
Q^{\prime\, 2}=-q^{\prime\, 2}_{max}=
\frac{(1-x_2)}{4}S_{ee} {\rm sin}^2\, \alpha_{max}.
\end{eqnarray}
The angle $\alpha_{max}$ in (\ref{Qpr}) is
determined by the acceptance of a lepton in the detector
(see, for instance, \cite{Diehl00}) and the value of
the c.m. energy of the $ee$ collision $\sqrt{S_{ee}}$ is
$91\, {\rm GeV}$ at LEP1 and $195\, {\rm GeV}$ at LEP2.

\noindent
In (\ref{xsec5}), the cross section for the subprocess can be calculated
directly~; we have
\begin{eqnarray}
\label{xsec6}
\frac{d\sigma_{e\gamma\to e\rho^0\rho^0}}
{dQ^2\, dW^2\, d{\rm cos}\theta\, d\phi}=
\frac{\alpha ^3}{16\pi}
\frac{\beta}{S_{e\gamma}^2}\,
\frac{1}{Q^2}
\Biggl(
1-\frac{2S_{e\gamma}(Q^2+W^2-S_{e\gamma})}
{(Q^2+W^2)^2}
\Biggr)
\left| A_{(+,+)}(\cos{\theta},W^2) \right|^2
\end{eqnarray}
where
\begin{eqnarray}
\label{App}
\left| A_{(+,+)}(\cos{\theta},W^2) \right|^2=
\left(
\left|  {\bf V}(\cos{\theta},W^2) \right|^2 +
\left|  {\bf A}(\cos{\theta},W^2) \right|^2
\right).
\end{eqnarray}
In (\ref{App}), the squared and $\rho$ meson polarizations summed
functions $|{\bf V}|^2$ and $|{\bf A}|^2$ read~:
\begin{eqnarray}
\label{V2}
&&|{\bf V}(\cos{\theta},W^2)|^2=
\frac{1}{4} P^{\alpha_1\alpha_2}(p_1)P^{\beta_1\beta_2}(p_2)
\nonumber\\
&&\sum_{i,\,q}e_{q}^{2} V^{(i)}_{\alpha_1\beta_1}\int dy_1 E_-(y_1)
H^{\rho\rho,\,V}_{i,\,q}(y_1,\cos{\theta},W^2)
\sum_{j,\,q}e_{q}^{2} V^{(j)}_{\alpha_2\beta_2}\int dy_2 E_-(y_2)
H^{\rho\rho,\,V}_{j,\, q}(y_2,\cos{\theta},W^2)
\\
\label{A2}
&&|{\bf A}(\cos{\theta},W^2)|^2=
\frac{1}{4}P^{\alpha_1\alpha_2}(p_1)P^{\beta_1\beta_2}(p_2)
\nonumber\\
&&\sum_{i,\, q}e_{q}^{2} A^{(i)}_{\alpha_1\beta_1}\int dy_1 E_+(y_1)
H^{\rho\rho,\,A}_{i,\, q}(y_1,\cos{\theta},W^2)
\sum_{j,\, q}e_{q}^{2} A^{(j)}_{\alpha_2\beta_2}\int dy_2 E_+(y_2)
H^{\rho\rho,\,A}_{j,\, q}(y_2,\cos{\theta},W^2),
\end{eqnarray}
where
\begin{eqnarray}
\label{Ppr}
P_{\alpha\beta}(p)=\sum_{\lambda}
e^{(\lambda)}_{\alpha} e^{*\,(\lambda)}_{\beta}=
-g_{\alpha\beta}+\frac{p_{\alpha}p_{\beta}}{m^2_{\rho}}.
\end{eqnarray}
As mentioned before, the expressions (\ref{V2}) and
(\ref{A2}) do not depend on the azimuth $\phi$.

\noindent
The helicity squared amplitude when
the integration over $\cos{\theta}$ is implemented
may be written as
\begin{eqnarray}
\label{F1}
F_{(+,+)}(W^2)=\int\limits_{0}^{1} d{\rm cos}\theta
\left| A_{(+,+)}(\cos{\theta},W^2) \right|^2,
\end{eqnarray}
and the cross section takes the form
\begin{eqnarray}
\label{xsec7}
&&\frac{d\sigma_{ee\to ee\rho^0\rho^0}}{dQ^2}=
\frac{25\alpha^4}{36\pi} \int\limits_{0}^{1}
dx_2 F_{WW}(x_{2})
\Biggl(
\frac{1}{ x_2^2 S_{ee}^2 Q^2}
\int\limits_{W^2_{min}}^{W^2_{max}} dW^2 \beta \,
F_{(+,+)}(W^2)-
\Biggr.
\nonumber\\
\Biggl.
&&\frac{2}{ x_2 S_{ee} Q^2}
\int\limits_{W^2_{min}}^{W^2_{max}} dW^2
\frac{\beta \, F_{(+,+)}(W^2)}
{Q^2+W^2}+
\frac{2}{Q^2}
\int\limits_{W^2_{min}}^{W^2_{max}} dW^2
\frac{\beta \, F_{(+,+)}(W^2)}
{(Q^2+W^2)^2}
\Biggr).
\end{eqnarray}
In (\ref{xsec7}), the magnitudes of $W^2_{min}$ and
$W^2_{max}$ are defined by the interval covering most of detected
two $\rho^0$ events, which in the L3 case at LEP \cite{L3Coll} is~:
\begin{eqnarray}
\label{interval}
1.21\, {\rm GeV}^2 < W^2 < 9.0\, {\rm GeV}^2.
\end{eqnarray}
Since the $\rho$ meson width is large, the lower
limit is a matter of convention, but it can be less than $4 m_\rho ^2$.
Notice, also,  that the integrated function $F_{(+,+)}$ is independent
of $Q^2$ up to logarithms.
Besides, the exact $W^2$ dependence of this
quantity remains unknown unless some modeling is used.
However, the mean value theorem gives the possibility
to reduce the three different integrals over $W^2$ in (\ref{xsec7}) to
one integration. Indeed, the mean value theorem
reads~:
\begin{eqnarray}
\label{mvt1}
&&\int\limits_{W^2_{min}}^{W^2_{max}} dW^2
\frac{\beta \, F_{(+,+)}(W^2)}
{Q^2+W^2}=\frac{1}{Q^2+ \langle W_{1}\rangle ^2}
\int\limits_{W^2_{min}}^{W^2_{max}} dW^2
\beta \, F_{(+,+)}(W^2),
\\
\label{mvt2}
&&\int\limits_{W^2_{min}}^{W^2_{max}} dW^2
\frac{\beta \, F_{(+,+)}(W^2)}
{(Q^2+W^2)^2}=
\frac{1}{(Q^2+\langle W_{2}\rangle ^2)^2}
\int\limits_{W^2_{min}}^{W^2_{max}} dW^2
\beta \, F_{(+,+)}(W^2)
\end{eqnarray}
with two phenomenological parameters
$\langle W_{1}\rangle $ and $\langle W_{2}\rangle$.
Let us now discuss the status of these parameters.
In principle, each of these parameters is a  function of $Q^2$,
but owing to the unknown $W^2$ dependence of hadronic function
$F_{(+,+)}$ the $Q^2$ dependences of given parameters stay out
of the exact computations. Therefore we will deal with
$\langle W_{1}\rangle $ and $\langle W_{2}\rangle$ which
will be considered in the sense of average value on the whole
interval of $Q^2$.
Notice also that the values of our parameters have actually
the same order of magnitude as $Q^2$, therefore we need to keep
$\langle W \rangle$-parameters in the prefactors of (\ref{mvt1})
and (\ref{mvt2}).

\noindent
>From (\ref{mvt1}) and (\ref{mvt2}), one can see that it is
useful to introduce a third phenomenogical parameter~:
\begin{eqnarray}
\label{C1}
C_1=\int\limits_{W^2_{min}}^{W^2_{max}} dW^2
\beta \, F_{(+,+)}(W^2).
\end{eqnarray}

\noindent
The normalization of our $2\rho$GDA's and  consequently the
value of our $C_1$ parameter (see, (\ref{sol_chi2})) is difficult to guess.
In the $\pi\pi$  case (see, for instance \cite{Diehl00}), the relation of the
second moment of the operator
defining the GDA to the energy momentum tensor,
allowed to relate the value of the second moment of the GDA to
the total energy carried by
quarks in the $\pi$ meson, which is known from the study of
parton distributions in the meson. This analysis required a modest
extrapolation of the meson-pair energy to zero, which was legitimate
in the two $\pi$ meson case with small $W$, but is very dangerous in
the present two $\rho$ meson case, due to the much larger threshold
$W_{min}$.
Still, one may expect, that some order of magnitude estimate
may be provided by such an extrapolation,
as will be confirmed
by the numerical results below. Moreover, such an analysis
may give indirect
access to the parton distributions inside a $\rho$ meson, and,
in particular, to the spin and orbital angular momenta carried by
quarks.

\noindent
Finally, the cross section (\ref{xsec7}) is  expressed through
these three phenomenological parameters in the following simple way:
\begin{eqnarray}
\label{xsec8}
&&\frac{d\sigma_{ee\to ee\rho^0\rho^0}}{dQ^2}=
\frac{25\alpha^4}{36\pi}\, C_1 \, \int\limits_{0}^{1} dx_2
F_{WW}(x_{2})
\Biggl(
\frac{1}{ x_2^2 S_{ee}^2 Q^2}+
\frac{2}{Q^2 (Q^2+\langle W_{2}\rangle ^2)^2}
-\frac{2}{ x_2 S_{ee} Q^2 (Q^2+\langle W_{1}\rangle ^2)}
\Biggr).
\nonumber\\
\end{eqnarray}
In the L3 Collaboration analysis,
the value $W$  belongs to the interval (\ref{interval}).
Hence we are able to conclude that the phenomenological parameters
$\langle W_{1} \rangle$ and $\langle W_{2}\rangle$ may take any values
inside this interval.

\section{Comparison of $\rho$-mesons and lepton pair productions}

\noindent
In this section we 
compare the $\rho^0\rho^0$ production  with the
production of a lepton pair $\mu^+\mu^-$ in the same kinematics.
The cross section of the process $e\gamma\to e \mu^+\mu^-$ 
coming through the $\gamma\gamma^*$ subprocess is
well-known and has the form :
\begin{eqnarray}
\label{lxsec}
\frac{d\sigma_{e\gamma\to e\mu\mu}^{\gamma\gamma^*}}{dQ^2}=
\frac{\alpha^3}{4 S_{e\gamma}^2\, Q^2}
\int\limits_{4m_{\mu}^2}^{W^2_{max}} dW^2 \Biggl(
1-\frac{2S_{e\gamma}(Q^2+W^2-S_{e\gamma})}
{(Q^2+W^2)^2}
\Biggr) f^{\mu^+\mu^-} (W^2),
\end{eqnarray}
where
\begin{eqnarray}
\label{fmuon}
f^{\mu^+\mu^-} (W^2)=8\biggl( {\rm ln}
\frac{1+\beta^{(\mu)}}{1-\beta^{(\mu)}}-
\beta^{(\mu)} \biggr), \quad \beta^{(\mu)}=\sqrt{1-\frac{4m_{\mu}^2}{W^2}} .
\end{eqnarray}
In (\ref{lxsec}), we can also apply the mean value theorem.
But now, by virtue of the known form of the muon function (\ref{fmuon}),
the  analogues of $\langle W_{i}\rangle$ can be explicitly
calculated. So, the mean value theorem for the muon case reads~:
\begin{eqnarray}
\label{mvt1mu}
&&\int\limits_{4m_{\mu}^2}^{W^2_{max}} dW^2
\frac{ f^{\mu^+\mu^-} (W^2)}
{Q^2+W^2}=\frac{1}{Q^2+ \langle W_{1,\,(\mu)}\rangle^2}
\int\limits_{4m_{\mu}^2}^{W^2_{max}} dW^2
 f^{\mu^+\mu^-}(W^2),
\\
\label{mvt2mu}
&&\int\limits_{4m_{\mu}^2}^{W^2_{max}} dW^2
\frac{ f^{\mu^+\mu^-}(W^2)}
{(Q^2+W^2)^2}=
\frac{1}{(Q^2+\langle W_{2,\,(\mu)}\rangle^2)^2}
\int\limits_{4m_{\mu}^2}^{W^2_{max}} dW^2
 f^{\mu^+\mu^-}(W^2).
\end{eqnarray}
Whence, we obtain
\begin{eqnarray}
\label{W1muon}
&&\langle W_{1,\,(\mu)} \rangle^2(Q^2)= R_1(Q^2)-Q^2,
\quad
\biggl(Q^2+\langle W_{2,\,(\mu)}\rangle^2\biggr)^2 = R_2(Q^2),
\end{eqnarray}
where the following notations are introduced~:
\begin{eqnarray}
\label{RKL}
R_n(Q^2)=\frac{{\cal K}}{{\cal L}_n(Q^2)} \quad
{\cal K}= \int\limits_{4m_{\mu}^2}^{W^2_{max}} dW^2
 f^{\mu^+\mu^-}(W^2), \quad
{\cal L}_n(Q^2)= \int\limits_{4m_{\mu}^2}^{W^2_{max}} dW^2
\frac{ f^{\mu^+\mu^-}(W^2)}{(Q^2+W^2)^n}.
\end{eqnarray}
The $Q^2$- dependences of $\langle W_{i,\,(\mu)}\rangle^2$ are shown
on Fig. 2. The solid line correspond to the function
$\langle W_{1,\,(\mu)}\rangle^2(Q^2)$ and the dashed one to the function
$\langle W_{2,\,(\mu)}\rangle^2(Q^2)$.
The weak $Q^2$- dependence of
$\langle W_{i,\,(\mu)}\rangle^2$ justifies the possibility to use the
 averaged $\langle W_{i,\,(\mu)}\rangle^2$ when fitting the data.
Besides, a fitting procedure gives us
the following representation for these functions~:
\begin{eqnarray}
&&\langle W_{1,\,(\mu)}\rangle^2(Q^2)=3.63-0.015\, Q^2+0.47\, {\rm ln}Q^2,
\nonumber\\
&&\langle W_{2,\,(\mu)}\rangle^2(Q^2)=2.90-0.024\, Q^2+0.73\, {\rm ln}Q^2 .
\nonumber
\end{eqnarray}

\noindent
Further,  the cross section of two $\rho$ meson
production  can be written as
\begin{eqnarray}
\frac{d\sigma_{ee\to ee\rho^0\rho^0}}{dQ^2}=
\int\limits_{0}^{1} dx_2  F_{WW}(x_{2})
\frac{\alpha^4}{4\pi}\frac{C_1}{Q^2 S_{e\gamma}^2}
{\cal N}(x_2)
\int\limits_{4 m_{\mu}^2}^{W^2_{max}} dW^2 \Biggl(
1-\frac{2S_{e\gamma}(Q^2+W^2-S_{e\gamma})}
{(Q^2+W^2)^2}
\Biggr) f^{\mu^+\mu^-}(W^2),
\end{eqnarray}
where the function ${\cal N}(x_2)$ is defined by
the ratio
( one reminds here that $S_{e\gamma}$ is proportionnal to the $x_2$
fraction, see (\ref{See})):
\begin{eqnarray}
\label{N}
&&{\cal N}(x_2)=\frac{{\cal I}_1(x_2)}{{\cal I}_2(x_2)},
\nonumber\\
&&{\cal I}_1(x_2)=\frac{25}{9}
 \Biggl(
1-\frac{2S_{e\gamma}}{Q^2+\langle W_1\rangle ^2}+
\frac{2S_{e\gamma}^2}
{(Q^2+\langle W_2 \rangle^2)^2}
\Biggr),
\nonumber\\
&&{\cal I}_2(x_2)= \Biggl(
1-\frac{2S_{e\gamma}}{Q^2+\langle W_{1,\,(\mu)}\rangle^2}+
\frac{2S_{e\gamma}^2}
{(Q^2+\langle W_{2,\,(\mu)}\rangle^2)^2} \Biggr)\, {\cal K}.
\end{eqnarray}
If one considers the case where $Q^2$ is large with respect  to
the invariant mass squared $W^2$,
we can omit the terms of $O(1/Q^2)$ and $O({\rm ln}Q^2/Q^4)$ in
(\ref{N}) and
obtain that ${\cal N}$ becomes independent of $x_2$~:
\begin{eqnarray}
\label{Nval}
{\cal N}\approx \frac{25}{9\, {\cal K}}= 0.01 \, {\rm GeV}^{-2}.
\end{eqnarray}

\noindent
We stress that this value has been obtained providing the
value of upper limit $W_{max}^2$ is fixed (see, (\ref{interval})).
Actually, the value
${\cal N}$ is a function of $W_{max}^2$, owing to the
$W_{max}^2$-dependence of the integral ${\cal K}$ (see, (\ref{RKL})),
and this dependence is pretty strong. For example, enhancing
$W_{max}^2$ up to $16.0\, {\rm GeV}^2$ the value ${\cal N}$ will
be halved.

\noindent
Thus, the cross section of our process is related to
the cross section of lepton pair production as
\begin{eqnarray}
\label{relat}
\frac{d\sigma_{ee\to ee\rho^0\rho^0}}{dQ^2}
\sim
0.01\,\, C_1 \,\, \frac{d\sigma_{ee\to e\mu\mu}^{\gamma\gamma ^*}}{dQ^2}.
\end{eqnarray}
In the next section we will show that  $C_1$ is close to
$1.0\, {\rm GeV}^2$.
Hence, one can see that the cross section of two $\rho$ meson production
is suppressed compared to the cross section of two muon production
by a factor which is approximately equal to $100$.
Note that a factor of suppression of the same order
was present in the case of two $\pi$ mesons production
\cite{Diehl00}.
\begin{figure}[htb]
%\vspace{9pt}
%\framebox[55mm]{\rule[-21mm]{0mm}{43mm}}
\includegraphics[width=16pc]{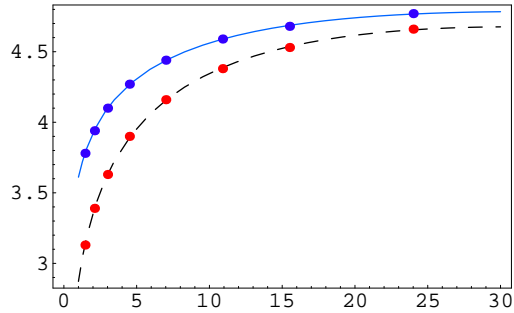}
\caption{ The muonic parameters $\langle W_{1,\,(\mu)}\rangle^2$
(solid) and $\langle W_{2,\,(\mu)}\rangle^2$ (dashed)
as a functions of $Q^2$.}
\end{figure}

\section{Comparision with experimental data}

\noindent
In the previous section we obtained a simple expression for
the two mesons cross section as a function of the three parameters 
$\langle W_{1}\rangle$, $\langle W_{2}\rangle$ and $C_1$.
Let us now make a fit of these phenomenological parameters
in order to get the best description of experimental data.
The best values of the parameters
can be found by the method of least squares, $\chi^2$-method,
which flows from the maximum likelihood theorem (see, for instance,
\cite{Mey}).
As usual, the $\chi^2$-sum as a function of parameters is written
in the form~:
\begin{eqnarray}
\label{chi2}
\chi^2=\sum_{i=1}^{N}\Biggl(
\frac{\sigma ^{exp}_{i} - \sigma ^{th}_i({\bf P})}{\delta\sigma_i}
\Biggr)^2,
\end{eqnarray}
where ${\bf P}=\{\langle W_{1}\rangle,\,\langle W_{2}\rangle,\, C_1\}$
denotes the set of fitted
parameters~; $\sigma ^{exp}_{i}$ and $\sigma ^{th}_i$ are
the experimental measurements of the cross section and its
theoretical estimations~; $\delta\sigma_i$ are the statistical errors.
The experimental data for the cross section of the exclusive
double $\rho^0$ production were taken from the measurement
of the L3 collaboration at LEP \cite{L3Coll}.

\noindent
Minimizing $\chi^2$-sum in (\ref{chi2}) with respect to the
parameters ${\bf P}$ we find that the set of solutions ${\bf P}_{min}$
with their confidence intervals are the following~:
\begin{eqnarray}
\label{sol_chi2}
C_1=1.20 \pm 0.23 \, {\rm GeV}^2 ,
\quad \langle W_1\rangle=3.0 \pm 1.9\, {\rm GeV},
\quad \langle W_2\rangle=1.50 \pm 0.09 \, {\rm GeV}.
\end{eqnarray}
With this the magnitude of $\chi^2$ is equal to $1.40$ and, therefore,
we have
\begin{eqnarray}
\frac{\chi^2}{{\rm degree\, of \, freedom}}=0.28 < 1
\end{eqnarray}
The confidence intervals were defined for the case of one-standard
deviation. Fig. 3 shows
the experimental data with the  theoretical fit.

\begin{figure}[htb]
%\vspace{9pt}
%\framebox[55mm]{\rule[-21mm]{0mm}{43mm}}
\includegraphics[width=16pc]{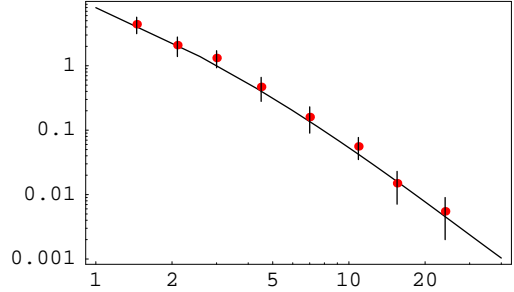}
\caption{Cross-section
$d\sigma_{ee\to ee\rho^0\rho^0}/dQ^2 [pb/GeV^2]$ as a function
of $Q^2$ .
The theoretical cross-section is plotted for the best fitted
parameters which are $C_1=1.2\, {\rm GeV}^2$ ,
$\langle W_{1}\rangle=3.0 \, {\rm GeV}$ and
$\langle W_{2}\rangle=1.5 \, {\rm GeV}$.}
\end{figure}

\noindent
We can see from  (\ref{sol_chi2}) that
the confidence interval for the parameter $\langle W_1\rangle$
covers the whole available interval for $W$.Moreover the obtained values of 
$\langle W_2 \rangle $ is quite small.
This analysis shows the compatibility of the data with the leading twist analysis which we
developed. At the same time, one cannot exclude the existence of sizeable higher
twist contributions to the production amplitude. More theoretical and experimental
studies are required to make a definite conclusion on  higher twist contributions.

\section{Conclusion}

\noindent
Data thus are in full agreement with the theoretical expectation
on the $Q^2$ behaviour of the cross section. Since
the effective structure function $F_{(+,+)}$ is expected
to be independent
of $Q^2$ up to logarithms,
our analysis of the data is a strong indication of the relevance
of the partonic
description of the process $\gamma\gamma ^*\to\rho ^0\rho ^0$
in the kinematics of the L3 experiment.

\noindent
Much more can be done if
detailed experimental results are collected. For instance the angular
dependence of the final state is a good test of the validity of the
asymptotic form of the generalized distribution
amplitudes. The spin structure of the final
state, if elucidated, would allow to disentangle the roles of the nine
generalized distribution amplitudes. The $W^2$ behaviour of the cross
section may have some interesting features. It depends much on the
possible resonances which are able to couple to two $\rho$ mesons.
The $\rho^+ \rho^-$ channel may be calculated along the same lines. In
that case a brehmstrahlung subprocess where the mesons are radiated from
the lepton line must be added. The charge asymmetry then comes from the
interference of the two processes. These items will be discussed in a
forthcoming publication.
More data may be collected at other energies,
in particular, in $e^+e^-$ experiments around $10 \, {\rm GeV}$
such as BABAR in SLAC and BELLE in KEK.

\section{Acknowledgements}

\noindent
We thank I. Boyko, M. Diehl, A. Nesterenko,
A. Olchevski and I. Vorobiev
for useful discussions and correspondance. This work has been
supported  in part
by RFFI Grant 03-02-16816 and by INTAS Grant (Project 587, call 2000).

\end{document}